\documentclass[structabstract]{aa}
\usepackage{graphicx}

\usepackage{epsf}
\usepackage{natbib}
\usepackage{amsmath}
\bibpunct{(}{)}{;}{a}{}{,}
\begin{document}

\def \jgr{J. Geophys. Res.}
\def \grl{Geophys. Res. Lett.}
\def \nat{Nature}
\def \mnras {MNRAS}
\def \aap {A{\&}A}
\def \aj {AJ}
\def \apj {ApJ}
\def \apjl {ApJL}
\def \gca {Geochim. Cosmochim. Acta}
\def \ssr {Space Sci. Rev.}
\def \asr {Adv. Space Res.}
\def \planss {Planet. Space Sci.}
\def \dth {\frac{\partial}{\partial\theta}}
\def \dfi {\frac{\partial}{\partial\phi}}
\newcommand{\eq}[1]{\begin{equation} #1 \end{equation}}
\newcommand{\bigz}[1]{\left( #1 \right)}
\newcommand{\bbigz}[1]{\left[ #1 \right]}
\newcommand{\bbbigz}[1]{\left\{ #1 \right\} }
\newcommand{\vek}[3]{\left(\begin{array}{c} #1 \\ #2 \\ #3 \end{array}
\right)}
\newcommand{\parderi}[2]{\frac{\partial #1}{\partial #2}}
\newcommand{\parderii}[2]{\frac{\partial^2 #1}{\partial #2^2}}
\newcommand{\deri}[2]{\frac{d #1}{d #2}}
\newcommand{\derii}[2]{\frac{d^2 #1}{d #2^2}}
\newcommand{\av} [1] {\langle #1 \rangle}
\newcommand{\ve}[1]{{\bf #1} }
\newcommand{\cnt}{\mathcal{C}}
\newcommand{\rsa}{\widetilde{\mathcal{C}}}
\newcommand{\Wmk}{Wm$^{-1}$K$^{-1}$}
\newcommand{\mj}{{$^{-1}$}}
\newcommand{\md}{{$^{-2}$}}
\newcommand{\mt}{{$^{-3}$}}
\newcommand{\ti}[1]{$\times 10^{#1}$}
\newcommand{\sumsum}[1]{\sum\limits_{n=#1}^\infty\sum\limits_{k=-n}^n}

%
%

   \title{Rotation of cometary meteoroids}

   \author{D. \v{C}apek\inst{1}}

   \institute{Astronomical Institute of the Academy of Sciences,
              Fri\v{c}ova 298,
              251 65 Ond\v{r}ejov,
              Czech Republic\\
              \email{capek@asu.cas.cz}
             }

   \date{Received ??? / Accepted ???}  

\abstract
{}
{
Rotation of meteoroids due to gas drag during the ejection from cometary nucleus has not been studied yet. The aim of this study is to estimate the rotational characteristics of meteoroids
after their release from a comet during normal activity. The results can serve as initial conditions for further analyses of subsequent evolution of rotation in the interplanetary space. 
}
{
Basic dependence of spin rate on ejection velocity and meteoroid size was determined analytically. A sophisticated numerical model was than applied to meteoroids ejected from 2P/Encke comet. The meteoroid shapes were approximated by polyhedrons with several thousands of surface elements, which have been determined by 3D laser scanning method of 36 Earth rock samples. These samples came from three distinct sets with different origin and shape characteristics such as surface roughness or angularity. Two types of gas-meteoroid interactions and three gas ejection models (leading to very different ejection velocities) were assumed. The rotational characteristics of ejected meteoroid population were obtained by numerical integration of equations of motion with random initial conditions and random shape selection.
}
{
It was proved, that the results do not depend on specific set of shape models and that they are applicable to (unknown) shapes of real meteoroids. A simple relationship between median of meteoroid spin frequencies $\bar{f}$ (Hz), ejection velocities  $v_{\rm ej}$ (m\,s$^{-1}$) and sizes $D$ (m) was determined. For diffuse reflection of gas molecules from meteoroid's surface it reads: $\bar{f}\simeq 2\times 10^{-3} v_{\rm ej} D^{-0.88}$, and for specular reflection of gas molecules from meteoroid's surface it is: $\bar{f}\simeq 5\times 10^{-3} v_{\rm ej} D^{-0.88}$. 
The distribution of spin frequencies is roughly normal in $\log$-scale and it is relatively wide; $2\sigma$-interval can be described as $(0.1, 10)\times \bar{f}$. Most of meteoroids are non-principal axis rotators. The median angle between angular momentum vector and spin vector is $12^\circ$. About $60\%$ of meteoroids rotate in long axis mode. Distribution of angular momentum vectors is not random. They are concentrated in the perpendicular direction with respect to the gas flow direction. These results were determined for 2P/Encke comet, but their validity is general.
Attention must be paid if the gravitation of nucleus plays an important role. 
}
{}


\maketitle

%
%
\section{Introduction}
From the observations of meteors and bolides, there are several phenomena suggesting that meteoroids rotate. 
(i) The light-curves of some bright meteors show quasi-periodic brightness variations \citep{Spurnyetal2007}. This phenomenon,
which is called {\em flickering}, is sometimes interpreted as a result of rotation of non-symmetric meteoroid 
\citep[e.g.][]{BeechBrown2000, Beech2001, SpurnyBorovicka2001, BeechIllingworthMurray2003}. The rotational origin of flickering has
however been doubted and other explanations, such as autofluctuating mechanism or triboelectric effects, were suggested 
\citep[e.g.][]{BabadzhanovKonovalova2004, Borovicka2006, SpurnyCeplecha2008, Spurnyetal2012}.
(ii) Periodic variations in velocity of Lost City bolide were also interpreted as a result of changing cross-section due to rotation \citep{Ceplecha1996, CeplechaRevelle2005}. 
(iii) Initial radius of meteor trains \citep{HawkesJones1978} and (iv) non-linear meteor trails \citep{Beech1988} can  also be a result of meteoroid rotation, 
as well as the meteoroid bursting in the atmosphere \citep[e.g.][]{StokanCampbellbrown2014}.
Unfortunately, precise and reliable determination of preatmospheric rotation from observations of meteors and bolides represents a significant problem so far.

The preatmospheric rotation of meteoroids (and more generally, evolution of rotation in interplanetary space) can be studied theoretically. For such studies, it is 
necessary to describe the action of the processes that may affect the rotation.
It was shown, that the radiative effects more efficiently affect the rotation of meteoroids in space than collisions with dust \citep{OlssonSteel1987}. 
The asymmetry parameter determined by \citet{Paddack1969} together with the time spent in interplanetary space have been used 
by many authors for estimates of the spin rate of meteoroids \citep{}, but detailed study which would describe 
the whole physics of meteoroid rotation in space self-consistently, is still missing.

For further modeling of the subsequent spin evolution in interplanetary space, the knowledge of initial rotation, just after the meteoroid birth, is also necessary.
For asteroidal meteoroids, which originate as debris from collisions in the Main Belt, the initial rotation can be estimated from results of hypervelocity
fragmentation experiments \citep[e.g.][]{Fujiwaraetal1989, Martellietal1994, GiblinFarinella1997, Giblinetal1998}.

The majority of shower meteoroids are released from parent cometary nucleus during the normal activity of the comet by gas drag 
\citep{Whipple1950, Whipple1951}. The gas drag mechanism is connected with sublimation of ice at the surface of the nucleus
and acceleration of embedded dust grains and pebbles by gas flow away from the comet.
If the meteoroid has irregular shape with some degree of windmill asymmetry, the gas may also
accelerate its rotation - similarly as in the simple experiment of \citet{Paddack1969}. 
Although many authors
dealt with the ejection process \citep[e.g.][]{Crifo1995, Jones1995, CrifoRodionov1997, Fulle1997, MaWilliamsChen2002, Molinaetal2008}, 
the rotation of meteoroids caused by gas drag during the ejection has not been studied yet. 

The aim of the present study is to fill this gap in our understanding of the meteoroids' rotation and to estimate the rotation characteristics of the meteoroids after the ejection from the parent cometary nucleus. It is the first in the assumed series of articles devoted to rotation of meteoroids. In Sec.~\ref{analSec} a simple analytical theory is described, Sec.~\ref{numSec} is devoted to the description of a sophisticated numerical model. The results from the numerical model can be found in Sec.~\ref{resSec}

\section{Simple analytical model}
\label{analSec}
The meteoroid motion and related acceleration of rotation during ejection can be described by the following simple analytical model.
Let us assume, that the cometary nucleus is spherical with radius $R_{\rm c}$ and mass $M_{\rm c}$. Sublimation of the cometary material due to solar heating causes a gas flow in the radial direction from the nucleus along coordinate $z$ ($z=0$ in the center of the nucleus). The meteoroid has mass $m$ and size $D$. Let us assume two forces acting on the meteoroid. The first one is gravitational force of the nucleus, which can be expressed as

\eq{F_{\rm g}=-GmM_{\rm c}/z^2, \label{eqgrav}} where $G=6.67\times 10^{-11}$\,N\,m$^2$\,kg\md is the gravitational constant. Molecules of the gas interact with meteoroid surface and cause drag force, which can be expressed in a simplified form as

\eq{F_{\rm gas}=A/z^2,} where $A$ is a positive constant \citep[e.g. review of][]{Ryabova2013}. The equation of motion is

\eq{m\deri{v}{t}=F_{\rm gas}+F_{\rm g},} where $v$ is the velocity in the $z$-direction. Using the identity $dv/dt = v\,dv/dz$ and initial conditions $v=0$ and $z=R_{\rm c}$ (meteoroid is lying at the surface of the nucleus), the dependence of the meteoroid velocity on distance $z$ can be found:

\eq{v=\sqrt{2\bigz{\frac{A}{m}-GM_{\rm c}}\bigz{\frac{1}{R_{\rm c}}-\frac{1}{z}}}.}
If the body has an amount of windmill asymmetry, the gas produces not only force $F_{\rm gas}$, but also torque $M_{\rm gas}$. Let us assume, that the torque is related to the drag force through effective moment arm $r_{\rm ef}$ \citep{Paddack1969} as 

\eq{M_{\rm gas} = r_{\rm ef} F_{\rm gas},} or with help of dimensionless asymmetry parameter $\xi=r_{\rm ef}/D$ \citep{PaddackRhee1975} as

\eq{M_{\rm gas} = \xi D F_{\rm gas}.} 
The equation of motion for rotation is 

\eq{C \deri{\omega}{t} = M_{\rm gas},} where $C$ is moment of inertia and $\omega$ is angular velocity. After substitution it has form:

\eq{C\deri{\omega}{z}=r_{\rm eq}A\bbigz{{2\bigz{\frac{A}{m}-GM_{\rm c}}}\bigz{\frac{1}{R_{\rm c}}-\frac{1}{z}}}^{-1/2}z^{-2}.} Then, the relationship between spin frequency and velocity of the spherical meteoroid (i.e. $C=8mD^2/5$) can be derived as:

\eq{f=\frac{5\xi}{\pi D}v\bigz{1-\frac{GM_{\rm c}m}{A}}^{-1}. \label{eqi}}
We can see that the spin frequency of the meteoroid is directly proportional to the velocity and inversely proportional to the body size. There is no explicit dependence on meteoroid density, but it is included in the velocity $v$ and also in the term in parentheses. The term in parentheses on the right hand side corresponds to the ratio of gravitational and drag force. If the gravitational force is small in comparison with the drag force, this term approaches zero. This is the case of small meteoroids\footnote{Gravitational force depends on size as $\propto D^3$, while drag force is proportional to $\propto D^2$.}. The term grows with increasing importance of the gravitational force (i.e. with increasing mass, size, or density of meteoroid). It is infinity when these two forces are equal. The explanation is as follows: The gravitational force does not affect the meteoroid rotation itself, but it reduces the meteoroid speed. The gas flow has therefore more time to accelerate the rotation. 

Eq.~(\ref{eqi}) was derived with an assumption that the rotation is continuously accelerated. In real situations the steady acceleration (or deceleration!) begins after a phase of chaotic evolution of rotation (see Sec.~\ref{individualSec}). Due to this fact, it is more useful to write Eq.~(\ref{eqi}) without the term in parentheses. In this case $\xi$ has the meaning of effective asymmetry parameter, which may slightly depend on size.

\section{Numerical model}
\label{numSec}

%
%
The analytical formula (\ref{eqi}) for spin frequency was derived with crude simplifications and assumptions concerning the meteoroid shape, force and torque expression and the motion.
Moreover, the analytical model is not able to predict the value of asymmetry parameter $\xi$, and it is not able to describe the distribution of spin frequencies, directions of the spin axes, degree of tumbling, etc.

In the following text, a more precise numerical model, that is not limited by above mentioned simplifications will be described. It uses meteoroid shapes approximated by polyhedrons which were obtained by 3D laser scanning of Earth rock samples. The force and torque of the gas flow is computed for each surface facet and then integrated over whole meteoroid surface. The rotational motion is computed by numerical integration of Euler's equations.

\subsection{Meteoroid shapes}
\label{shapeSec}
\begin{figure}
\centering
        \begin{tabular}{cc}
            \includegraphics[width=4cm]{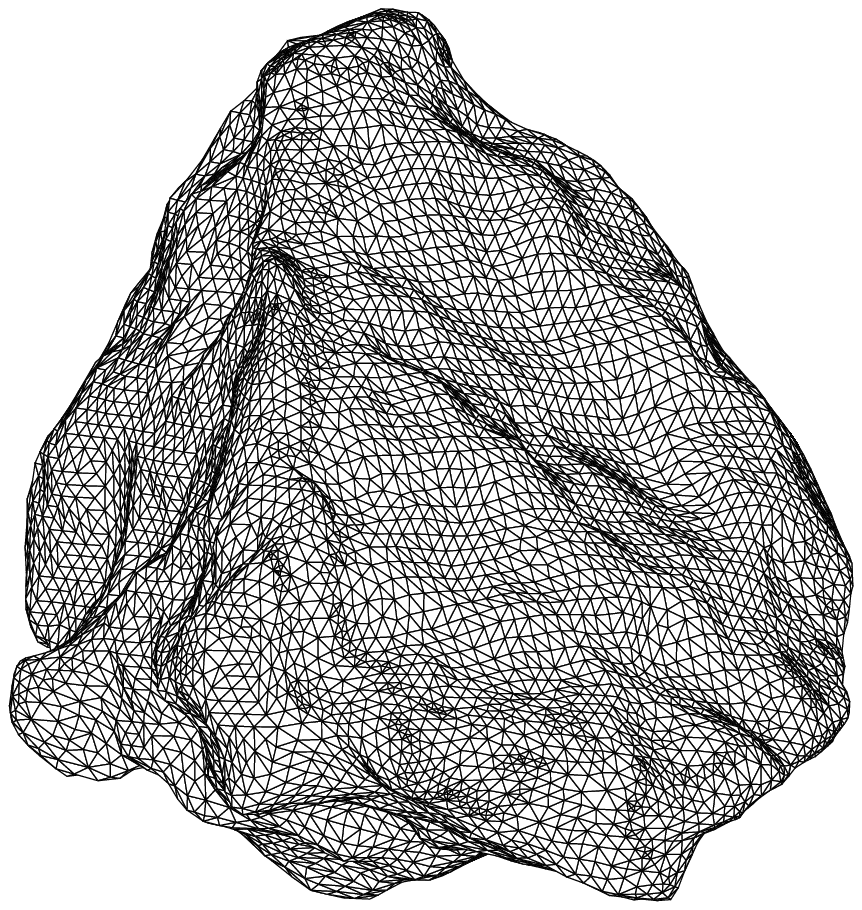}&
            \includegraphics[width=4cm]{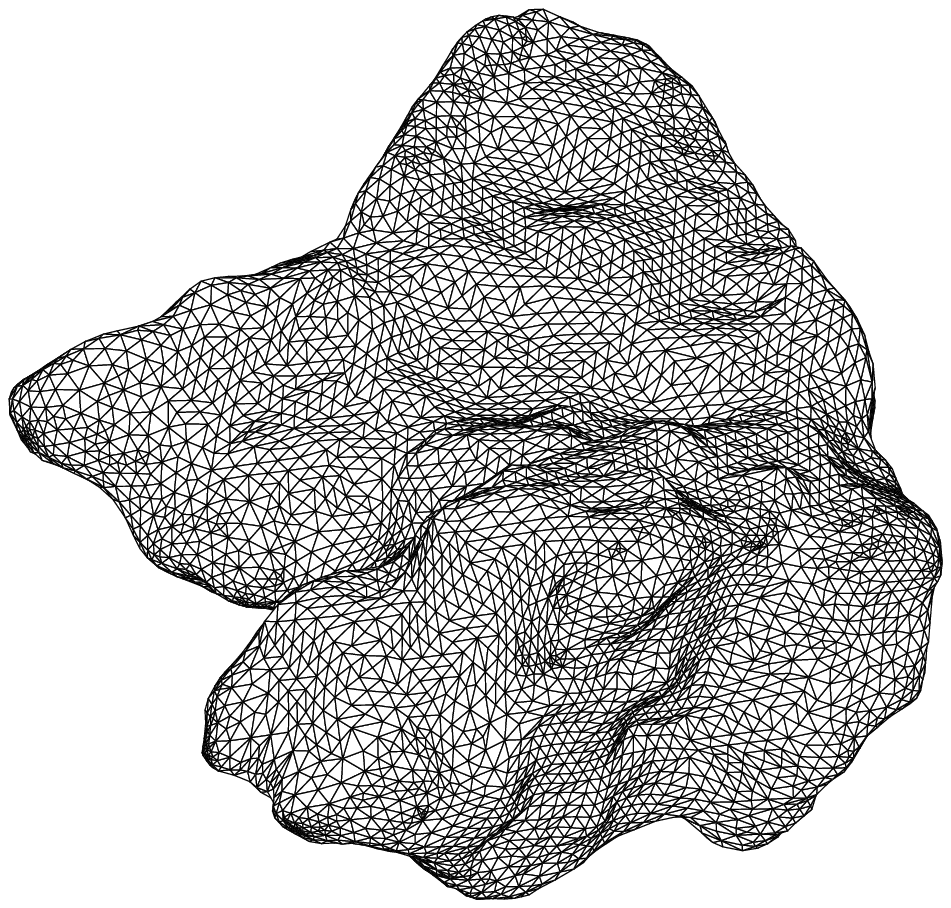}\\
            \includegraphics[width=4cm]{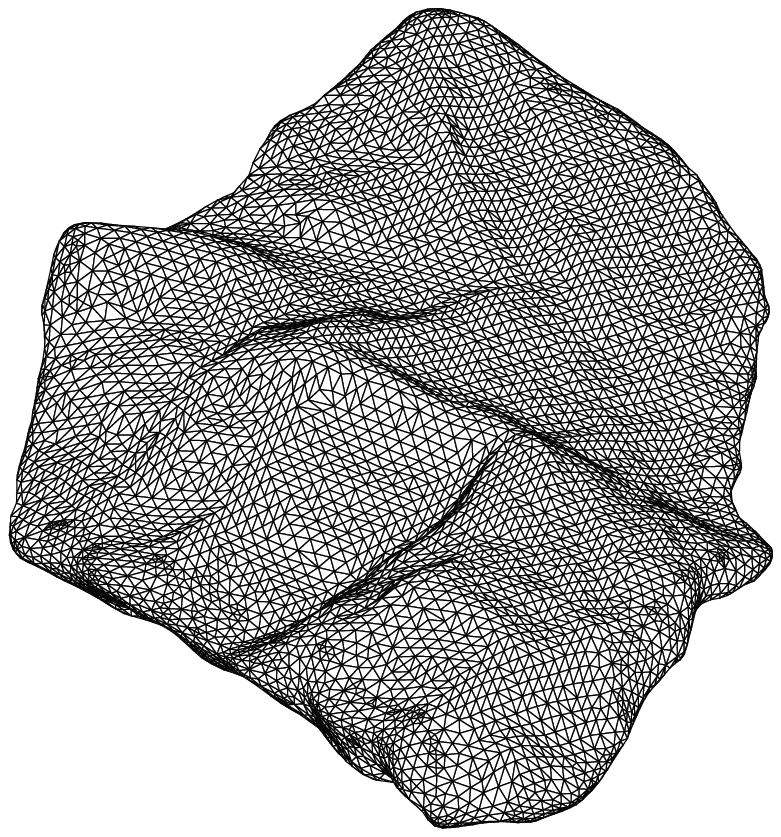}&
            \includegraphics[width=4cm]{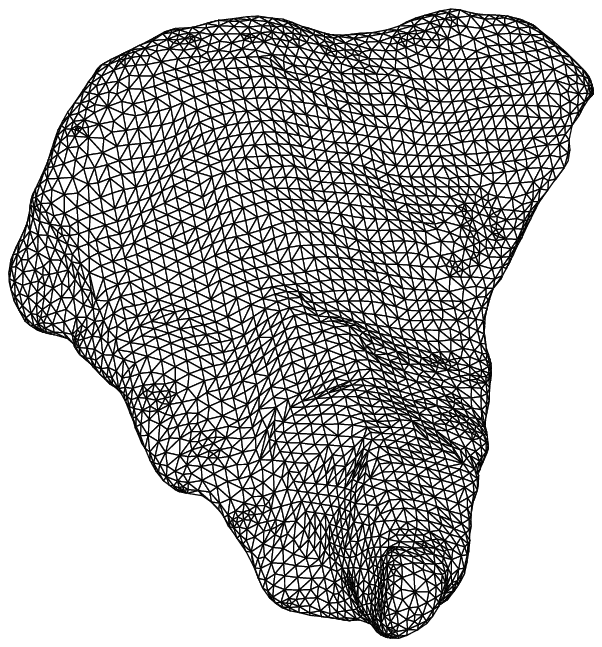}\\
            \includegraphics[width=4cm]{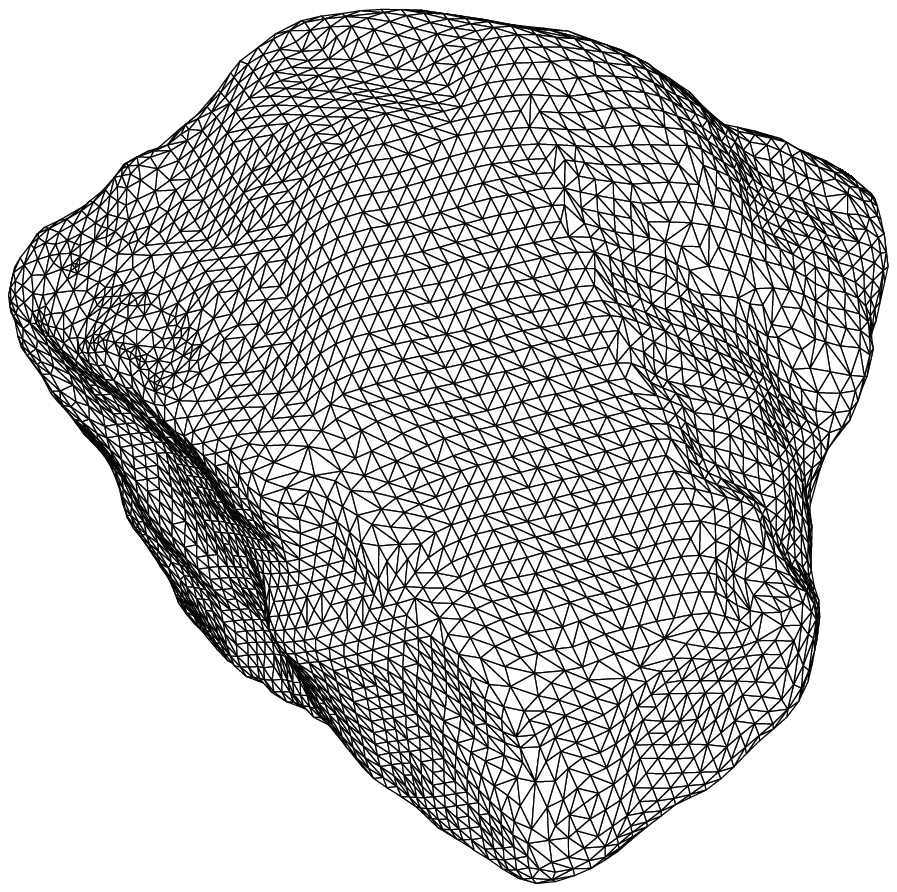}&
            \includegraphics[width=4cm]{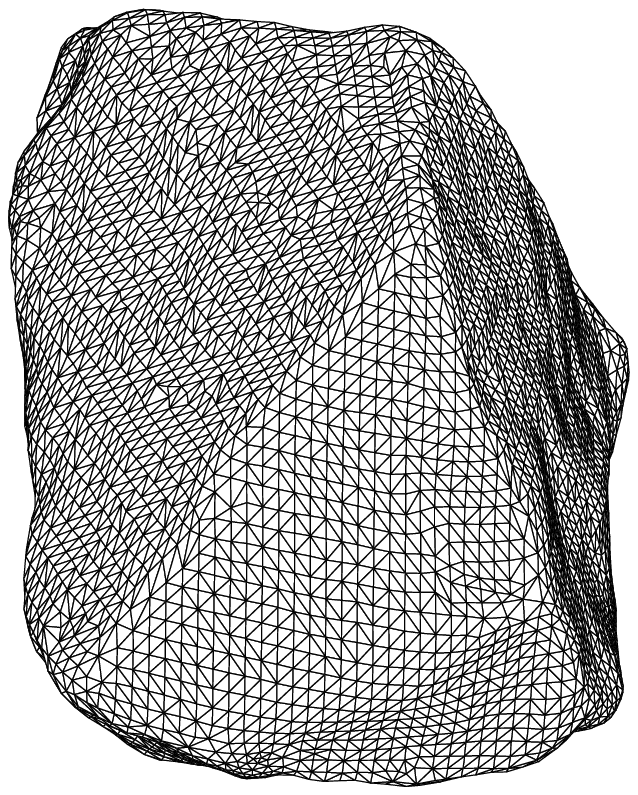}\\
        \end{tabular}
\caption{Example of the polyhedral models for meteoroid shapes. They were obtained by 3D laser scanning of 36 samples of various terrestrial rocks. 
Upper row: clay, middle row: trachybasalt fragments, lower row: gravel. The plots are not to scale.}
\label{shapesFig}
\end{figure} 

The shape of a body represents one of the most important quantities because it controls the ability
of the object to be spun up. There is no torque on symmetrical bodies like spheres, cubes, three-axial ellipsoids 
or blocks. The meteoroid must have a certain amount of windmill asymmetry which causes that 
the gas flow will be able to spin up the body. 

What do the shapes of cometary meteoroids look like? It represents one of the main difficulties in this 
modeling. I decided to digitize the terrestrial samples which differ in their origin, shape characteristics
and strength, and test how the results depend on the sample origin. For this purpose I chose three different 
sets of samples. Each set consists from 12 shapes which were selected randomly from larger group of samples to avoid preferential selection of
``nice'' shapes by a collector.

The first set is composed from fragments of a volcanic rock (trachybasalt) which was broken apart by a hammer. These shapes 
are usually planar and sharp and I call them ``fragments''. 
The second set consists from pieces of gravel (a metamorphic rock). These samples are more round than fragments 
and they are referred to as ``gravel''. 
The third set contains pieces of broken block of dry clay. They have very bumpy surface, sometimes with holes and open 
cracks. They are called ``clay''.

All shapes were digitized by 3D laser scanning method by SolidVision~s.r.o. company. The precision of the shape determination
is better than 0.5\,mm and the resulting shapes are represented by polyhedrons with several thousands of triangular facets (Fig.~\ref{shapesFig}). The  volume of samples ranges from 0.51 to 17.29\,cm$^3$. The number of surface elements depends of the surface area and it ranges from 3\,548 to 35\,242 facets.

The polyhedral description allows a simple determination of volume, mass, tensor of inertia of the body as well as the surface normal, area and radiusvector to the center of each surface facet. These quantities can be used for determination of the forces and the torques acting on the meteoroid.

\subsection{Gas ejection}
\label{gasejSec}
The ejection of meteoroids during normal cometary activity is caused by the gas flow, which escapes from the cometary nucleus due to sublimation of its material. The gas density $\rho_{\rm gas}$ and velocity $v_{\rm gas}$ represent important quantities which determine the magnitude of the force. These quantities were studied by many authors dealing with theoretical estimates of ejection velocity. It is obvious that different models give different results for ejection velocities \citep[e.g. review of][]{Ryabova2013}. Therefore, three distinct models were chosen for comparison: 

The model {\sf J1995-100} is based on \cite{Jones1995}. Assuming spherical symmetric ejection, the gas density was determined from Eqs.~(10)--(12) in \cite{Jones1995}. The gas velocity was computed from Eqs.~ (3), (4), (8) and those for gas density in \cite{Jones1995}. In this model, the adiabatic expansion of the gas is assumed, which causes that the gas velocity increases with distance $z$ from the nucleus, and the gas density has a more complicated dependence than $z^{-2}$. 

Next two models are based on \cite{MaWilliamsChen2002}. In this case the gas velocity $v_{\rm gas}$ is constant and equal to velocity of water molecules at sublimation temperature, which is 580\,m\,s\md. The gas density can be expressed from Eqs.~(4) and (24) in \cite{MaWilliamsChen2002} as: 

\eq{\rho_{\rm gas}(z) = \frac{1}{4\alpha} \frac{S_\odot}{H}\bigz{\frac{R_{\rm c}}{z}}^2\bigz{\frac{1}{r^2}-\frac{1}{r_{\rm s}^2}}\frac{1}{\bar{v}_{\rm gas}},} where $\alpha$ is the fraction of the nucleus surface area that is active, $S_\odot$ is the Solar constant, $H$ sublimation heat, $r$ heliocentric distance, and $r_{\rm s}$ heliocentric distance of the beginning of cometary material sublimation. Model {\sf M2002-050} assumes ejection from sunlit hemisphere (i.e. $\alpha=0.5$), and {\sf M2002-002} assumes that only $2\%$ of the surface is active. 

Thus the three gas ejection models result in large range of the gas action magnitude. Model {\sf J1995-100} corresponds to weak gas flow, since the ejection from whole cometary nucleus surface (including night side) is assumed. The other extreme is model {\sf M2002-002}, which assumes that the whole mass is ejected in a very narrow jet, the result of which is very strong gas flow. 
Model {\sf M2002-050} represents a conservative case. 

\subsection{Force, torque and meteoroid motion}
Gas molecules interact with the surface of the meteoroid and cause drag force and torque. The present model assumes free-molecular flow regime. It means that the size of meteoroids is smaller than the mean free path of molecules and the interaction of gas flow with meteoroid can be described as impacts of solitary molecules. (This assumption is however not fully met in case of larger meteoroids, small heliocentric distances and proximity to the surface of the nucleus.)  Two possible types meteoroid-gas interaction are assumed: (i) specular reflection which corresponds to the ideally elastic collisions, and (ii) diffuse reflection which means temporary capture and emission of gas molecules in a random direction. In case of the specular reflection, the force acting on $i$-th small surface facet is:

\eq{ d\ve{f}_i=\rho_{\rm gas} (v_{\rm gas}-v)^2\,\bbigz{-2(\ve{n}_i\cdot\ve{e})^2 \ve{n}_i} dS_i,} where
$\ve{n}_i$ is unit outer normal to the surface, $\ve{e}$ is unit vector in the direction of the gas flow, $dS_i$ is the area of the surface facet. This formula can be simply expressed in case of polyhedral description of meteoroid shapes.  For the diffuse reflection:

\eq{d\ve{f}_i=\rho_{\rm gas} (v_{\rm gas}-v)^2\, (\ve{n}_i\cdot\ve{e}) \bigz{\ve{e}- \frac{2}{3}\ve{n}_i} dS_i.} Total force caused by gas is given by a sum over the whole surface:

\eq{\ve{F}_{\rm gas}=\sum\limits_i d\ve{f}_i.} The gravitational force also affects the meteoroid's motion and it is given by Eq.~(\ref{eqgrav}). The centrifugal force is neglected in the model. The total torque reads

\eq{\ve{M}_{\rm gas}=\sum\limits_i \ve{r}_i\times d\ve{f}_i,} where $\ve{r}_i$ is radiusvector to the center of the $i-$th surface facet. The torque is caused only by gas flow, because gravitational torques are negligible for such small bodies.

The translational and rotational motions were computed simultaneously, since the total force and torque depends both on the orientation and distance from the nucleus (and also on the velocity). 
The motion of meteoroid can be described by radiusvector $\ve{r}$, velocity $\ve{v}$, three Euler's angles $\varphi$, $\vartheta$, $\psi$ and three components of the angular velocity $\bf{\omega}$. 
For the numerical integrations of the Euler's equations are, however, more suitable Euler's parameters $q_0, q_1, q_2, q_3$, because they don't have singularities in the poles \citep[e.g.][]{Fukushima2008}\footnote{Note the typographical error in Eq.~A31 of \citet{Fukushima2008}, where a factor of 2 is missing in the expression of $\theta$ angle.}. 
The system of equations of motion for translation and rotation was solved numerically by 4th order Runge-Kutta method with variable timestep. The suitability of the numerical method was proved by conservation of energy and angular momentum tests.

%

%
%

\subsection{Model parameters and computation details}
\label{paramsSec}
The numerical model was applied to 2P/Encke comet, which belongs to the Taurid complex. 
The nucleus of this comet has mean effective radius $R_{\rm c}=3950\pm60$\,m \citep{LowryWeissman2007} and mass $M_{\rm c}=9.2\pm5.8\times 10^{13}$\,kg \citep{SosaFernandez2009}. The perihelion distance is 0.33\,AU and semimajor axis is 2.2178\,AU. The meteoroid bulk densities are assumed to be the same as for Taurid meteoroids $\rho=1.6$\,g\,cm\mt \citep{BabadzhanovKokhirova2009, Madiedoetal2014}. Three sizes of meteoroids, corresponding to equivalent spheres of diameters $1$\,mm, $1$\,cm and $10$\,cm were studied.
The computational scheme was following:
\begin{itemize}
\item Sizes of all meteoroid shapes were rescaled so that they have the same volume as the sphere of assumed size (1\,mm, 1\,cm or 10\,cm).
\item Position on the orbit was randomly selected and appropriate amount of meteoroids was released. The amount was proportional to the mass loss rate (i.e. $\propto\rho_{\rm gas} v_{\rm gas}$).
\item In the beginning of the integration, the shape and the initial orientation of each meteoroid was selected randomly. The integration started from the surface of the nucleus with zero velocity.
\item The equations of rotational and translational motion were integrated. If the meteoroid fell back to the surface, it was rejected. (This case was more frequent for large meteoroids at the larger heliocentric distances.) Othervise, the integration stopped when it reached the distance of $25\times R_{\rm c}$.
\item The values of ejection velocity, spin frequency, direction of the moment of inertia, degree of tumbling, etc. were saved.
\end{itemize}
In total, $\sim$800\,000 integrations were done, which took $\sim700$ CPU days.

\section{Results}
\label{resSec}

%
%

\subsection{Rotation of individual meteoroids}
\label{individualSec}
\begin{figure}
\centering
        \begin{tabular}{c}
            \includegraphics[width=8cm]{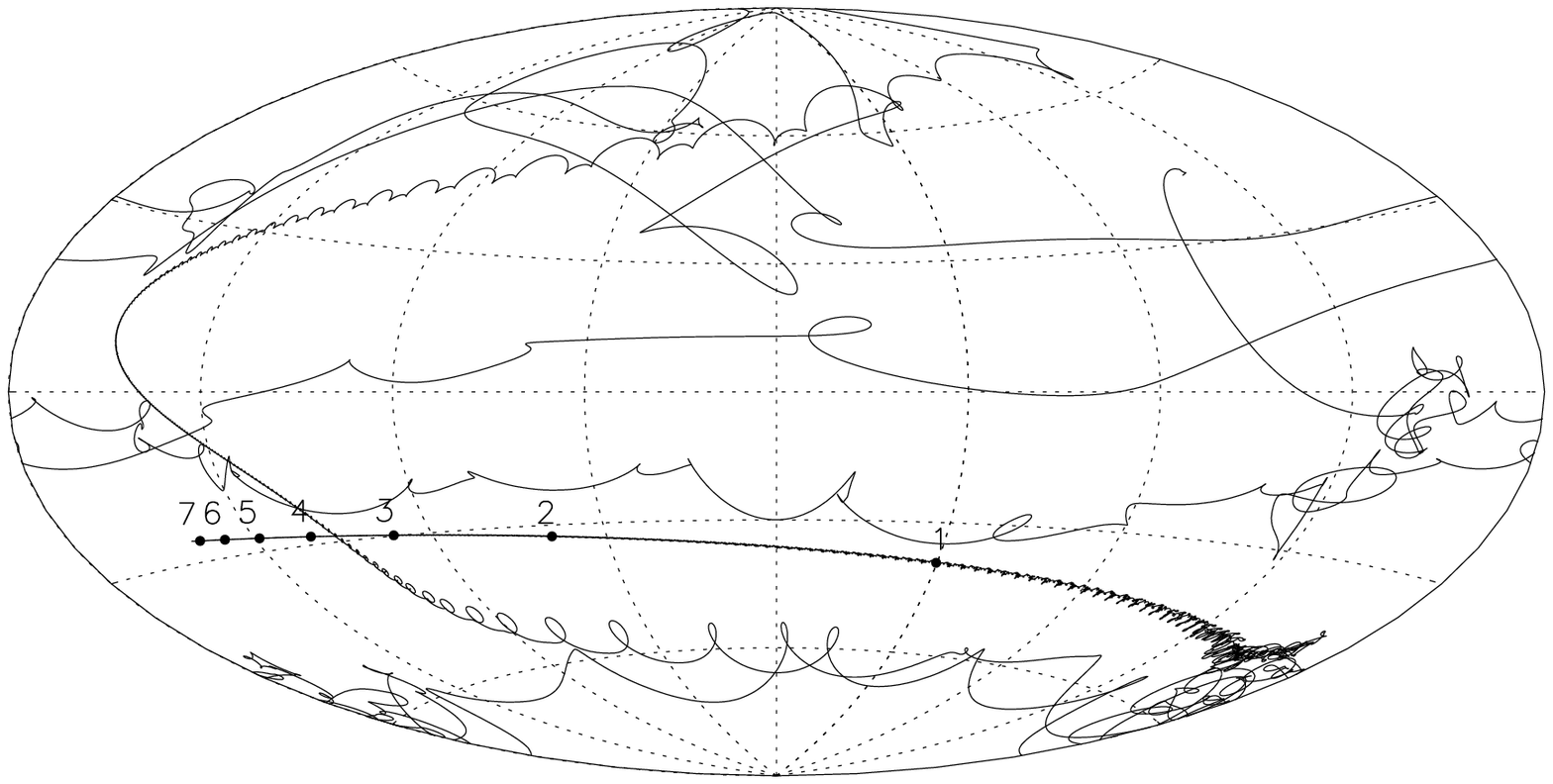}\\
            \includegraphics[width=8cm]{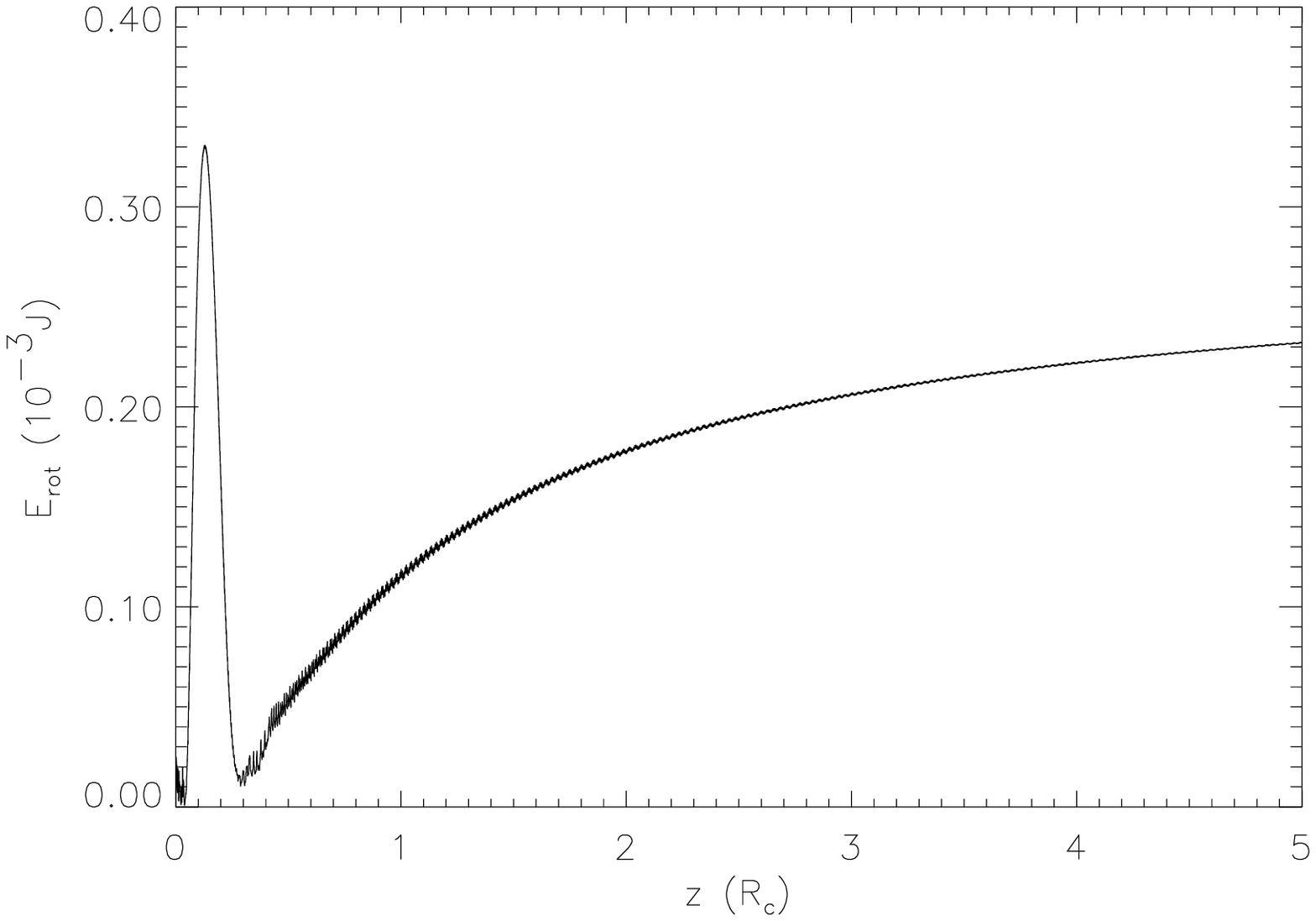}\\
        \end{tabular}

\caption{Example of evolution of rotation of 1\,cm meteoroid. Upper plot: Evolution of angular momentum direction with respect to the inertial frame. The numbers denote the distance from the surface in terms of nucleus radii $R_{\rm c}$. Lower plot: Evolution of the rotational part of the kinetic energy. The stage I of chaotic rotation and stage II of regular rotation can be clearly distinguished in both plots. In this case the transition between them occurs at about $0.3\times R_{\rm c}$ above the surface.}
\label{exampleFig}
\end{figure} 

The main aim of this study is to estimate the rotational properties of meteoroids far from the nucleus. It is however useful to briefly describe the evolution of rotation for individual meteoroids. The rotation during the ejection process can be divided into two stages: 
\paragraph{I. chaotic rotation.} After the release from the surface, the meteoroid usually rotates chaotically. It wobbles and tumbles, the spin axis orientation and direction of angular momentum changes in random way and the rotation speed is alternately accelerated and decelerated (Fig.~\ref{exampleFig}). 
\paragraph{II. regular rotation.} After some time, the meteoroid rotation begins to evolve more regularly. The direction of angular momentum slowly drifts to a final stage, spin axis moves about it and the rotation speed is monotonically accelerated or decelerated (Fig.~\ref{exampleFig}). With increasing distance from the nucleus, the gas density decreases as well as the gas forces and torques acting on the body. Far from the nucleus, the meteoroid has a constant velocity, constant angular momentum vector and kinetic energy. 

The transition between both stages is not sharp and it is difficult to estimate typical time necessary for the transition from the stage I to stage II. The transition is usually reached at the distance from the nucleus, where the gas density is still sufficiently high. The steady evolution of rotation therefore begins not at surface but at this distance, whereas the velocity is accelerated from the surface. Moreover, the spin rate at the beginning of the stage II is not zero and it can be both accelerated and decelerated. This results in a deviation of frequency-velocity dependence from the Eq.~(\ref{eqi}). Due to higher inertia and reduced velocity of larger meteoroids caused by a gravitational force, the stage of regular rotation begins at lower heights than for smaller meteoroids. Although the larger bodies rotate more slowly than the smaller ones, finally they have higher ratio of rotational to translational part of the kinetic energy. 

The ejection from the surface to the distance of $25\times R_{\rm c}$ lasted roughly from 10 minutes for $1$\,mm bodies, to 3 hours for 10\,cm meteoroids in perihelion.

%
%

\subsection{Spin frequency}
\begin{figure*}
\centering
        \begin{tabular}{ccc}
            \includegraphics[width=5.8cm]{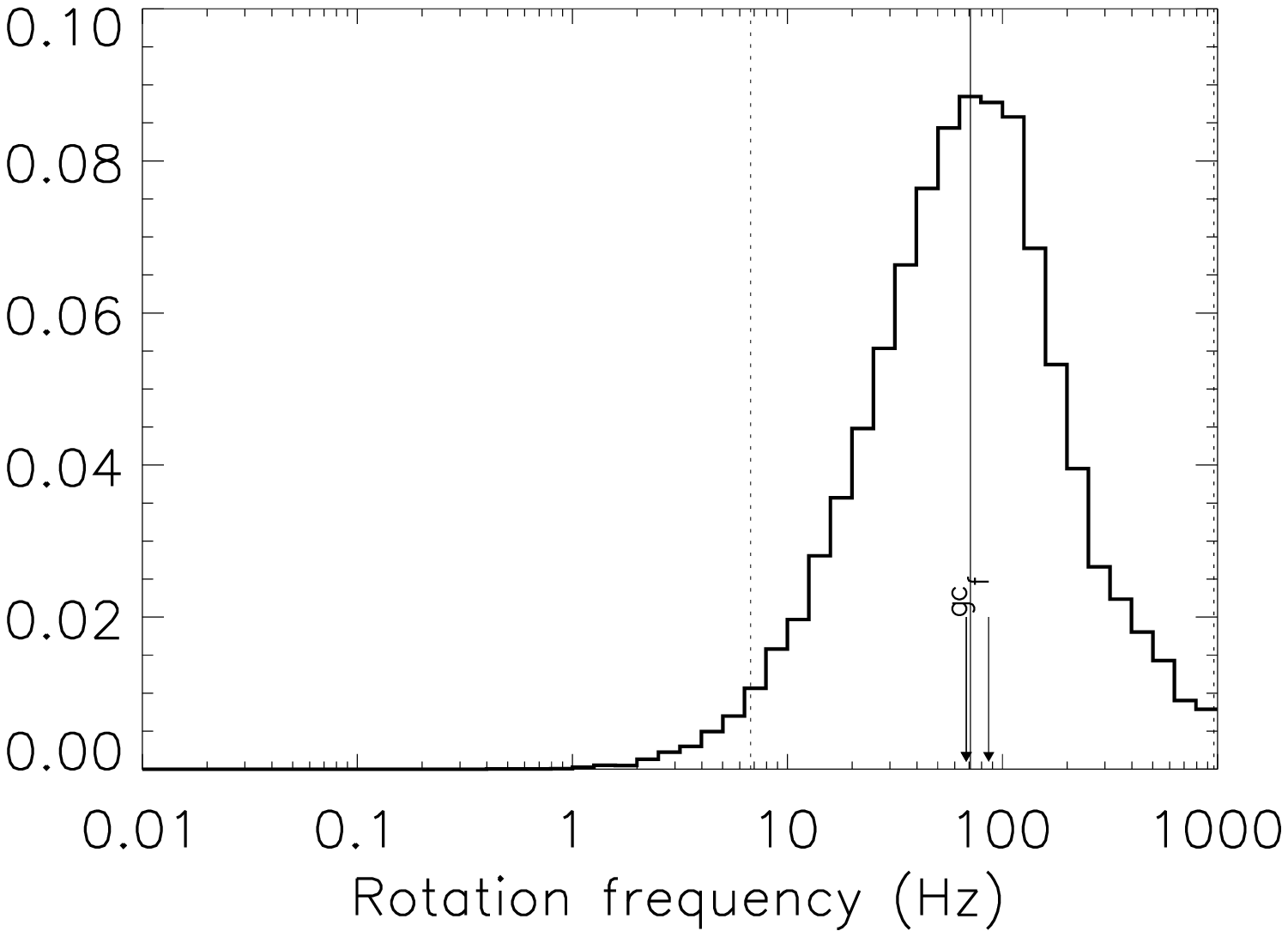}&
            \includegraphics[width=5.8cm]{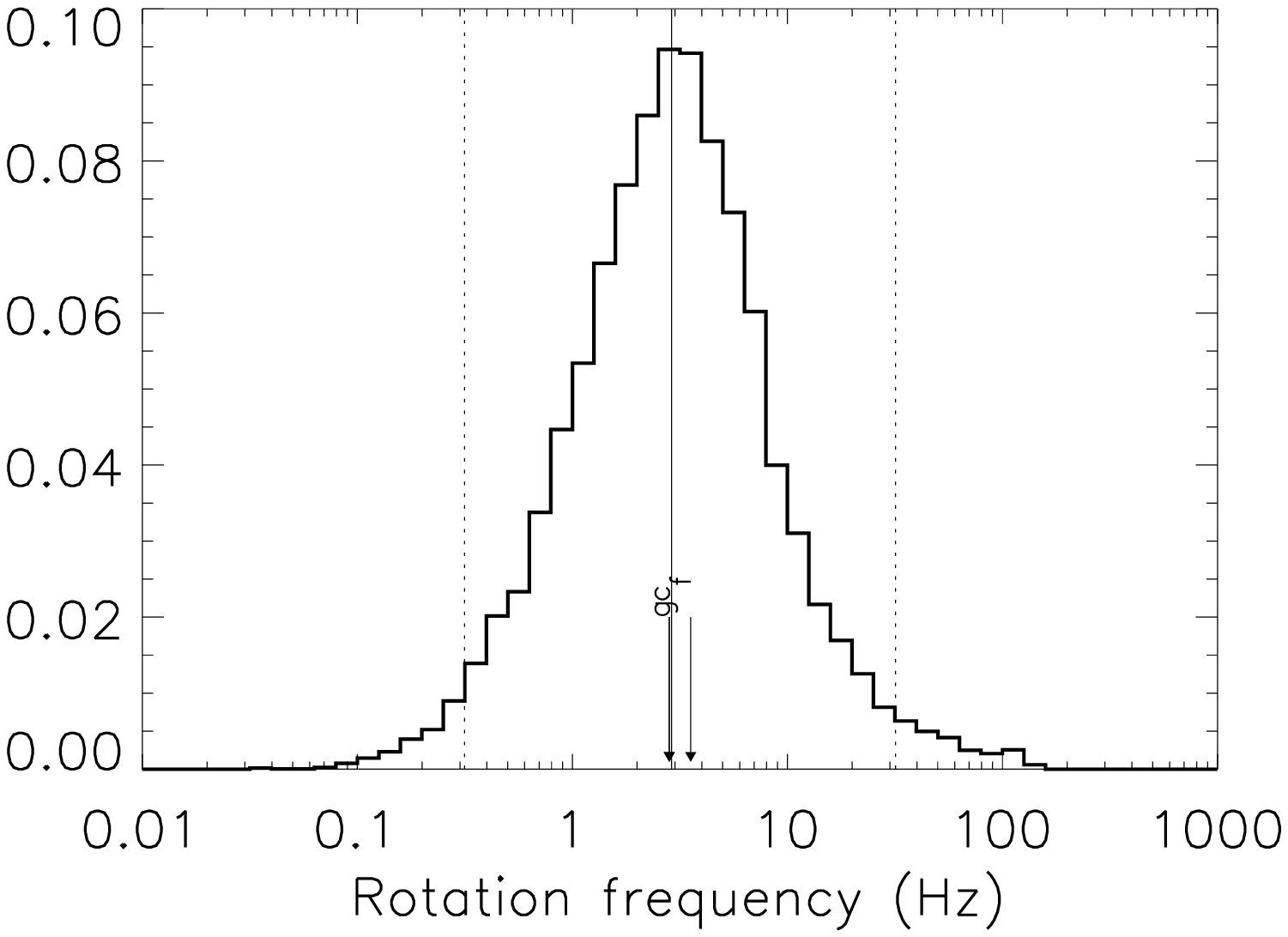}&
            \includegraphics[width=5.8cm]{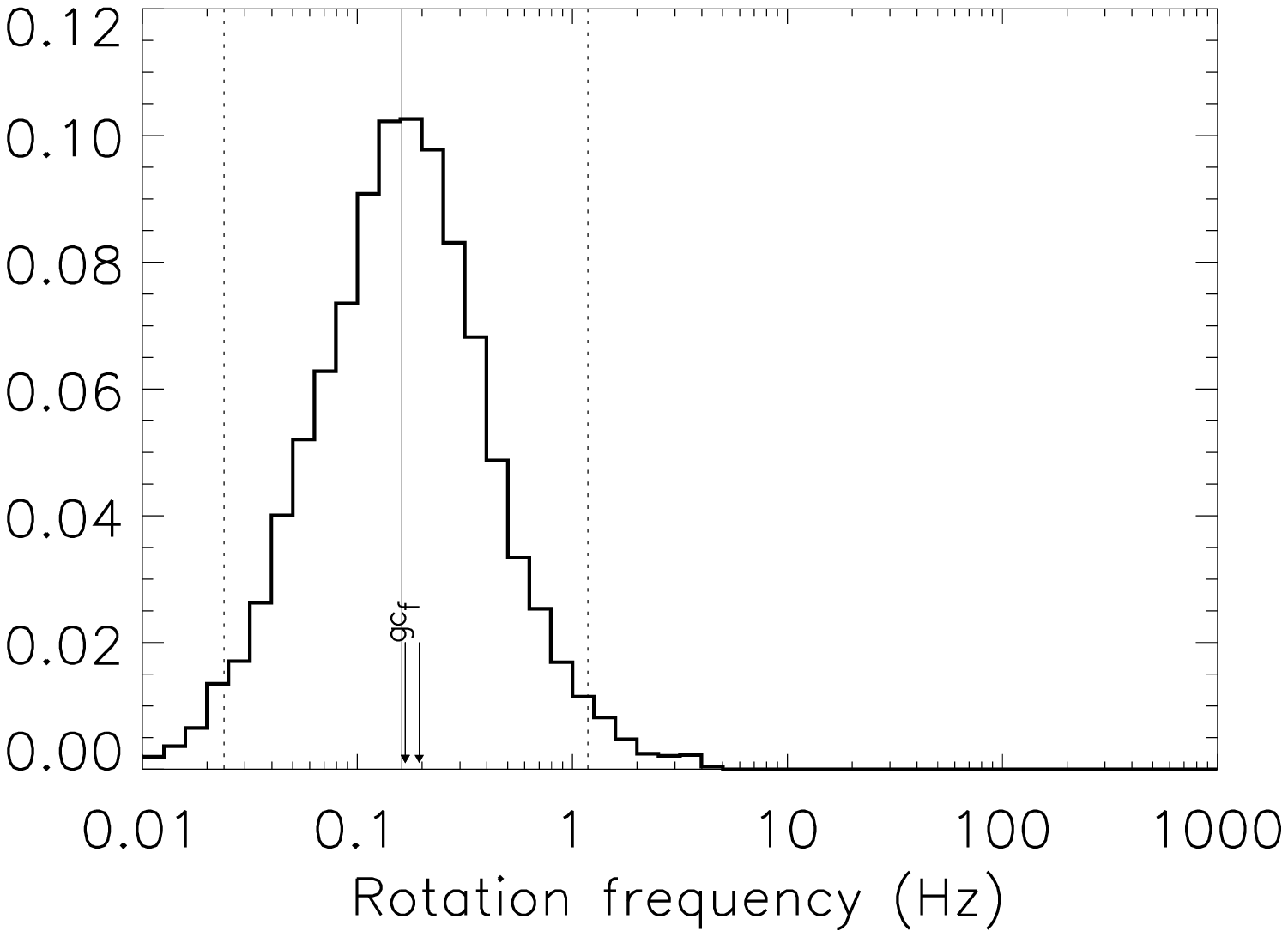}\\
        \end{tabular}
\caption{Distribution of spin frequencies after ejection from 2P/Encke comet for 1-mm (left), 1-cm (middle) and 10-cm (right) meteoroids. Gas ejection is according to
\citet{MaWilliamsChen2002}, active fraction of the surface is $50\%$. The solid vertical line represents median value and dotted vertical lines bound $2\sigma$ interval. Small arrows with letters {\it g}, {\it c}, {\it f} denote median values for gravel, clay and fragments shapes.}
\label{freqFig}
\end{figure*} 
The spin frequencies of meteoroids far from the nucleus were computed for three gas ejection models, two types of gas-meteoroid interactions and three sizes. Thus, 18 different distributions of spin frequencies were obtained. An example of the distribution of $f$ for three meteoroid sizes can be seen in Fig.~\ref{freqFig}. In this case, gas ejection model {\sf M2002-050} and diffuse reflection of gas molecules from the surface of meteoroids were considered. The meteoroid shapes of all three shape sets were used. The distribution is approximately normal in $\log$ scale and its width can be expressed by boundary values of $2\sigma$ interval. Approximately $95\%$ of values lay inside of this interval.  For 1\,mm meteoroids, the median value of spin frequency is $\bar{f}=70.9$\,Hz, with $2\sigma$ interval (6.7-963)\,Hz. In case of 1\,cm bodies, median value $\bar{f}$ is 2.9\,Hz, with $2\sigma$ interval (0.3-31.8)\,Hz, and largest, 10\,cm, bodies have median $\bar{f}=0.16$\,Hz and $2\sigma$ interval (0.02-1.2)\,Hz. It can be seen that the spin frequencies of meteoroids with particular size lay inside relatively wide interval ranging across about one order of magnitude. The median frequency decreases with increasing size as $D^{-1.31}$ in this size range. Median frequencies determined separately for each shape set are almost the same. The difference between largest and smallest values is lower than $30\%$. It is negligible with respect to the width of $2\sigma$ intervals. 

\begin{table*}[t]
    \begin{center}
        \begin{tabular}{|r|c|rrrrrc|}
            \hline
ejection & $D$  & $\bar{f}_{\rm min}$ & $ \bar{f} $ & $\bar{f}_{\rm max}$ & $v_{\rm ej}$ & $\xi$            & $\Delta f_{\rm cfg}$\\
 model   & (mm) &      (Hz)     &  (Hz) & (Hz)          &    (m/s)     & $\times 10^{-4}$ & ($\%$)\\
			\hline
{\tt M2002-002 D} &   1 &   32.02 &  258.32 & 3002.13 & 345.12 &   4.70 &    30\\
{\tt M2002-050 D} &   1 &    6.74 &   70.92 &  962.90 & 101.93 &   4.37 &    27\\
{\tt J1995-100 D} &   1 &    4.81 &   49.16 &  729.80 &  71.86 &   4.30 &    30\\
{\tt M2002-002 D} &  10 &    1.50 &   12.93 &  124.64 & 139.54 &   5.82 &    26\\
{\tt M2002-050 D} &  10 &    0.31 &    2.89 &   31.84 &  31.14 &   5.83 &    26\\
{\tt J1995-100 D} &  10 &    0.21 &    1.90 &   22.48 &  22.18 &   5.37 &    29\\
{\tt M2002-002 D} & 100 &    0.08 &    0.69 &    5.12 &  48.94 &   8.92 &    18\\
{\tt M2002-050 D} & 100 &    0.02 &    0.16 &    1.18 &  11.82 &   8.56 &    16\\
{\tt J1995-100 D} & 100 &    0.02 &    0.12 &    0.89 &   8.55 &   8.45 &    23\\
{\tt M2002-002 S} &   1 &   64.77 &  556.67 & 4888.90 & 276.85 &  12.63 &     2\\
{\tt M2002-050 S} &   1 &   17.47 &  171.46 & 1732.48 &  89.24 &  12.07 &     7\\
{\tt J1995-100 S} &   1 &   11.97 &  124.24 & 1273.32 &  66.73 &  11.70 &    11\\
{\tt M2002-002 S} &  10 &    3.07 &   28.72 &  246.04 & 127.99 &  14.10 &     4\\
{\tt M2002-050 S} &  10 &    0.65 &    6.66 &   63.86 &  29.95 &  13.97 &     4\\
{\tt J1995-100 S} &  10 &    0.43 &    3.93 &   42.27 &  17.90 &  13.80 &     3\\
{\tt M2002-002 S} & 100 &    0.14 &    1.30 &   10.48 &  45.61 &  17.88 &     6\\
{\tt M2002-050 S} & 100 &    0.05 &    0.31 &    2.33 &  10.04 &  19.46 &    10\\
{\tt J1995-100 S} & 100 &    0.04 &    0.22 &    1.73 &   7.03 &  19.84 &     8\\
%
            \hline
        \end{tabular}
    \end{center}
        \caption{Resulting rotational characteristics of meteoroids ejected from 2P/Encke comet for meteoroid sizes $D$ 1\,mm, 1\,cm and 10\,cm, three ejection models (see Sec.~\ref{gasejSec}) and two types 
				of gas-meteoroid interactions ({\sf D} means diffuse and {\sf S} specular reflection of gas molecules from meteoroid's surface). 
				$\bar{f}$ is median of spin frequencies, bounds of $2\sigma$ interval are $\bar{f}_{\rm min}$ and $\bar{f}_{\rm max}$, $v_{\rm ej}$ mean ejection velocity,
				$\xi$ is asymmetry parameter and $\Delta f_{\rm cfg}$ represents maximum difference of $\bar{f}$ between the shape sets (see the text).}
        \label{resultsTab}
\end{table*}

The resulting medians of spin frequencies, $2\sigma$ intervals, median ejection velocities and other quantities for various gas ejection models and meteoroid sizes can be seen in Tab.~\ref{resultsTab}.
The median spin frequency depens on (i) meteoroid size, (ii) type of gas-meteoroid interaction and (iii) also on the gas ejection model. The dependence (iii) is caused by different ejection velocities corresponding to these models. The dependence of $\bar{f}$ on gas ejection model can be removed if asymmetry parameter $\xi$ and median ejection velocity $v_{\rm ej}$ is taken into account. The rotation of meteoroid evolves in a more complicated way than assumed in Sec.~\ref{analSec}, as was described in previous section. After a stage of chaotic acceleration and deceleration  of rotation and chaotic movement of angular momentum direction, the meteoroid reaches quasi-stable rotation, which is further uniformly evolved. Due to this fact, it is useful to determine the asymmetry parameter $\xi$ as

\begin{figure}
	\centering
    \includegraphics[width=8cm]{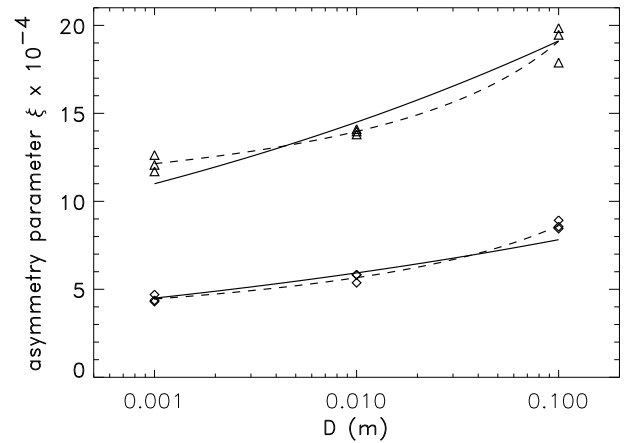}
	\caption{The dependency of asymmetry parameter $\xi$ on meteoroid size for three gas ejection models (see Sec.~\ref{gasejSec}) and two types of gas-meteoroid interaction. The diamonds correspond to numerical results for diffuse reflection and the triangles the specular reflection of gas molecules from meteoroid's surface. The solid and dashed lines represent interpolation of the size dependency (see the text).}
\label{xivsdFig}
\end{figure} 

\eq{\xi=\frac{\pi}{5}\,D\,\frac{\bar{f}}{v_{\rm ej}}, \label{xidefEq}}
where $v_{\rm ej}$ is the median of ejection velocity. The resulting values of $\xi$ can be seen in Tab.~\ref{resultsTab} and also in Fig.~\ref{xivsdFig}. The asymmetry parameter $\xi$ is almost independent on gas ejection model. It still depends on type of gas-meteoroid interaction and the size. Gas molecules specularly reflected from meteoroid's surface are able to spin up the meteoroid more quickly than molecules which are reflected diffusively. The ratio is about $2.5\times$. The dependency on size is caused by more complicated evolution of rotation during meteoroid ejection. It is related to the height of transition between stage I  and stage II which depends on meteoroid size (Sec.~\ref{individualSec}). The asymmetry parameter can be approximated as

\eq{\xi=\xi_0 \bigz{\frac{D}{D_0}}^{0.12},\label{xiEq}}
where $D_0=0.001$\,m, and $\xi_0=4.5\times 10^{-4}$ for diffuse reflection and $\xi_0=11\times 10^{-4}$ for specular reflection of gas molecules from the meteoroid's surface (solid line in Fig.~\ref{xivsdFig}). Obviously, better approximation can be found, e.g. rational function $\xi=1.55\,(\log D-6.73)/(\log D-0.39)$ for diffuse reflection and $\xi=8.28\,(\log D-1.56)/(\log D -0.11)$ for specular reflection of gas molecules from meteoroid (dashed line in Fig.~\ref{xivsdFig}). The differences between numerically determined medians $\xi$ and those from Eq.~(\ref{xiEq}) are however lower than $\sim 13\%$ which is substantially smaller than typical width of the distribution, so this relationship is sufficiently precise. The resulting relation for median spin frequency can be determined from (\ref{xidefEq}) and (\ref{xiEq}):

\eq{\bar{f}=\frac{5}{\pi}\xi_0\frac{v_{\rm ej}}{D}\bigz{\frac{D}{D_0}}^{0.12}. \label{resultEq}}
This equation enables to estimate the median of spin frequency of meteoroids ejected from comet 2P/Encke according to the size and ejection velocity. The meteoroid density is not explicitly present, but it determines the ejection velocity. The ejection velocity may be determined with more appropriate gas ejection model, active nucleus surface ratio or bulk density of meteoroids, or it may be determined from observations.


\subsubsection{Width of the distribution} As can be seen in Fig.~\ref{freqFig}, the spin frequency distribution is relatively wide. The width is a result of (i) different abilities of meteoroid shapes to be spun up, (ii) random initial orientation of meteoroid shapes and (iii) different heliocentric distances of the ejection. The lower and upper boundary of $2\sigma$ intervals, were determined for all distributions. The lower boundary vary between $(0.10-0.18)\times \bar{f}$ and the upper boundary between $(7.3-13.6)\times \bar{f}$. These limits may be roughly estimated as $0.1\times\bar{f}$ and $10\times\bar{f}$.


\subsubsection{Rotational bursting} The rotation of smaller meteoroids can be accelerated up to several hundreds or even thousands of Hertz. Is it possible that these bodies can be destroyed by centrifugal forces? The tensile stress due to centrifugal force inside rotating meteoroid can be estimated from a formula for the stress in the center of rotating sphere \citep[e.g.][]{Kadishetal2005}:

\eq{\sigma=\frac{\pi^2}{2}\rho D^2 f^2.}
After substitution from Tab.~\ref{resultsTab}, the stress increases with decreasing meteoroid size. Assuming gas ejection model {\sf M2002-050} and specular reflection of gas molecules, the stress for median spin frequency is $\sim 200$\,kPa for 1\,mm meteoroids, $\sim 35$\,Pa for 1\,cm meteoroids and $\sim 8$\,Pa for 10\,cm meteoroids. Due to wide distribution of the spin frequencies, some meteoroids can reach much stronger stresses. For example if spin frequency of 1\,mm meteoroids is $f=1000$\,Hz, the tensile stress due to centrifugal force is $\sim 8$\,kPa, which is comparable with the apparent strength of Taurid meteoroids \citep{TrigoLlorca2006}. Moreover the strength distribution of Taurid meteoroids shows large spread \citep{Brownetal2013} and therefore rotational bursting during ejection process can be expected for some of the fast small meteoroids.


\subsubsection{Differences between shape sets}
The shape of body is very important quantity, which determines the final spin state. Since there is no possibility how to obtain shapes of real cometary meteoroids, 36 shapes derived from Earth rock samples were used throughout the numerical modeling. But how good is the approximation of real shapes by these ones? The shapes belong to three sets according to the origin (clay, fragment, gravel - see Sec.~\ref{shapeSec}). Bodies of these sets differ in surface character (bumpy or smooth) and in overall shape (rounded or sharp). A simple test is how the results change for each shape set. It is expressed by column denoted as $\Delta f_{cfg}$ in Tab.~\ref{resultsTab}, which means $\max(\bar{f}_{\rm c},\bar{f}_{\rm f},\bar{f}_{\rm g})/\min(\bar{f}_{\rm c},\bar{f}_{\rm f},\bar{f}_{\rm g})$, where $\bar{f}_{\rm c}$, $\bar{f}_{\rm f}$, $\bar{f}_{\rm g}$ are medians of spin frequencies for clay, fragments and gravel shapes. It can be seen, that the results differs by less than $30\%$ for diffusive reflection of gas molecules and by less than $11\%$ for specular reflection of gas molecules from meteoroid surface. This is very good match. It indicates that the difference between results obtained by using shapes based on Earth rock samples and the results for potential real meteoroid shapes can be expected to be the same -- say within a factor of 2.

%
%

\subsection{Degree of tumbling}
\begin{figure}
	\centering
    \includegraphics[width=7.75	cm]{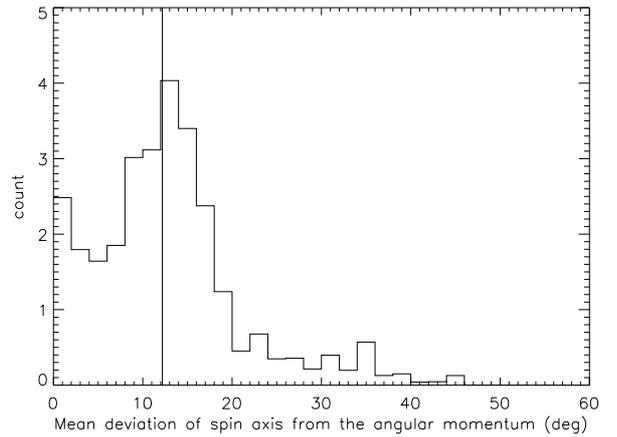}
	\caption{The distribution of mean angle between angular momentum vector and spin axis. Solid vertical line represents median, which is $12.3^\circ$.}
\label{L_W_angleFig}
\end{figure}

Another important information concerning the rotation of cometary meteoroids is, if they are principal axis rotators or non-principal axis rotators. Far from the cometary nucleus, where the torque due to escaping gas flow in negligible, the vector of angular momentum is constant. The spin axis of non-principal axis rotators rotates about the angular momentum vector along an unclosed trajectory. Mean angle between angular momentum vector and spin axis can be used as a degree of tumbling. (Another possibility is to use an angle between angular momentum vector and shortest axis of inertia tensor \citep[e.g.][]{Pravecetal2014}.) If it is zero, the body rotates about principal axis of the inertia tensor. The non-principal axis rotation can take place in small-axis mode or long-axis mode. In the first case, the spin axis moves about the body axis which corresponds to the largest moment of inertia, $I_3$. In the latter case, the spin axis moves about the body axis which corresponds to the smallest moment of inertia, $I_1$ \citep[e.g.][]{Pravecetal2005}. The distribution of mean angle between angular momentum vector and spin axis can be seen in Fig.~\ref{L_W_angleFig}. This distribution does not depend on the meteoroid ejection model. The median is $\sim 12^\circ$. Interestingly, in $\sim 60\%$ of cases, the rotation is in long-axis mode. This is partially a result of the meteoroid shape models, which are more elongated than flattened. 

%
%

\subsection{Angular momentum direction}
\begin{figure}
	\centering
    \includegraphics[width=8cm]{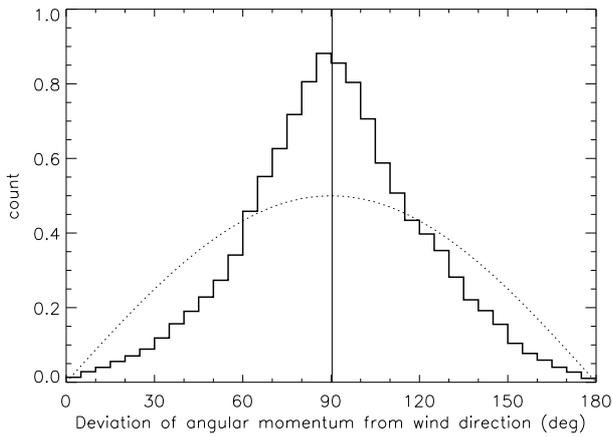}
	\caption{The distribution of deviation of the angular momentum vector from the direction of the gas flow direction. Sold vertical line represents median, which is $90.5^\circ$. Dotted curve corresponds to a random distribution of angular momentum vector directions.}
\label{L_Z_angleFig}
\end{figure}

The last quantity which was investigated is a direction of angular momentum vector. It was found, that its distribution is not random. The angular momentum vectors are concentrated towards the perpendicular direction with respect to the gas flow direction (see Fig.~\ref{L_Z_angleFig}). If it is real, it may help to explain some features of polarimetric observations of comets \citep[e.g. review of][]{Kolokolovaetal2004}.

\subsection{General validity of the results}
The relationship for median spin frequency (\ref{resultEq}) should be valid in general and it should be applicable also to bodies with different parameters than those in Sec.~\ref{paramsSec}. To check this assumption, the median spin rates were numerically determined for another radius and mass\footnote{The mass follows from assumption that the nucleus density is 1\,g\,cm\mt.} of 2P/Encke comet, $R_{\rm c}=2400$\,m \citep{Fernandezetal2000}, $M_{\rm c}=5.8\times 10^{13}$\,kg, Taurid meteoroid density $2.5$\,g\,cm\mt and compared with (\ref{resultEq}). Assuming gas ejection model {\sf M2002-050} and diffusive reflection, the numerical model leads to median spin frequency and ejection velocity of $46.54$\,Hz and $61.24$\,m\,s{\mj} for 1\,mm meteoroids, $2.11$\,Hz and $22.36$\,m\,s{\mj} for 1\,cm meteoroids and $0.12$\,Hz and $7.39$\,m\,s{\mj} for 10\,cm meteoroids. These values differ by $15\%$ from the results determined by (\ref{resultEq}). 

Similar computations were also performed for Perseid meteoroids. Gas ejection model {\sf M2002-050}, mass $1.2\times 10^{13}$\,kg and radius $1800$\,m of parent comet 55P/Tempel-Tuttle \citep{Jewitt2004} and density $0.4$\,g\,cm{\mt} of meteoroids were assumed. The resulting median frequencies from numerical modeling differs from those determined by Eq.~(\ref{resultEq}) by less than $15\%$. The results of the numerical modeling of Taurid meteoroids rotation can be thus cautiously applied to other meteoroid streams, which are the result of normal cometary activity. Some deviations from the results may occur when gravitation of cometary nucleus plays more important role than in the studied case. The rotation of meteoroids of various meteoroid streams will be studied in detail in a following paper.

\section{Discussion}
The present model is not able to determine specific value of meteoroid spin frequency after ejection from 2P/Encke comet due to the lack of reliable ejection velocity data. The dependence of median spin frequency on ejection velocity and meteoroid size (\ref{resultEq}) is however common for all three gas ejection model, despite of very different ejection velocities. The reliable estimate of the spin frequency therefore depends on reliable value of ejection velocity. Equation (\ref{resultEq}) can be rewritten into more simple form as

\eq{\bar{f}\simeq 2\times 10^{-3} v_{\rm ej} D^{-0.88}}
for diffuse reflection of gas molecules, and 
\eq{\bar{f}\simeq 5\times 10^{-3} v_{\rm ej} D^{-0.88}}
for specular reflection of gas molecules from meteoroid's surface ($\bar{f}$ in Hz, $v_{\rm ej}$ in m\,s\mj, and $D$ in m).
But the direct usage of (\ref{resultEq}) for estimates of preatmospheric spin rate of meteoroids is doubtful. During the time, which meteoroid spent in the interplanetary space, the rotation is affected by several phenomena. In the studied size range, the most important are radiative effects \citep{OlssonSteel1987}, i.e. windmill effect and YORP. The timescale of YORP evolution can be estimated by re-scaling of mean doubling time $t_{\rm d}=14$\,Myr, which was determined by \citet{CapekVokrouhlicky2004} for 2-km gaussian random spheres with spin period of 6\,hours on circular orbit with semimajor axis 2.5\,AU, assuming principal axis rotation in asymptotic states. Corresponding values are $\sim 4$ years for $1$\,mm (50\,Hz) and $70$ years for $10$\,cm (0.1\,Hz) Taurid meteoroids. The actual timescales will be however longer due to (i) heat diffusion through the volume of such small bodies \citep{Breiteretal2010}, (ii) evolution of the spin axis direction by YORP effect and (iii) non-principal axis rotation of the most of meteoroids. In any case, the preatmospheric rotation may correspond to the initial rotational state only in for short time spent in the interplanetary space, but this subject is beyond the scope of this article and it will be studied in the future.


\section{Summary}
\begin{itemize}
	\item Simple formula was derived for median of spin frequencies $\bar{f}$ (Hz) of meteoroids after ejection from a comet. It depends on meteoroid size $D$\,(m) and the ejection velocity $v_{\rm ej}$\,(m\,s$^{-1}$) as: $$\bar{f}\simeq 2\times 10^{-3} v_{\rm ej} D^{-0.88}$$ for diffuse reflection of gas molecules from meteoroid's surface, and $$\bar{f}\simeq 5\times 10^{-3} v_{\rm ej} D^{-0.88}$$  for specular reflection of gas molecules from meteoroid's surface. These formulae were determined for 2P/Encke, but they are generally valid and with caution they can be used for another comets. The dependence of median of spin frequencies on meteoroid density and on physical properties of cometary nucleus is hidden in the value of $v_{\rm ej}$.
	\item The distribution of spin frequencies is roughly normal in the $\log$-scale. It is relatively wide; more than 95$\%$ of values are inside an interval $(0.1,\; 10)\times \bar{f}$.
	\item Most of meteoroids are non-principal axis rotators. Median of mean angle between angular momentum vector and spin axes is $\sim 12^\circ$. 
	\item Angular momentum vectors are not distributed randomly in the space. They are concentrated at the perpendicular directions with respect to the gas flow.
	\item Meteoroid shapes were approximated by shape models derived from three distinct sets of different Earth rock samples. The results for these sets differ by less than 30$\%$ despite of different origin and shape characteristics of these sets. Therefore, the results are probably applicable to unknown shapes of real cometary meteoroids.
\end{itemize}

\begin{acknowledgements}
I thank to Petr Pravec and Pavel Spurn\'{y} for helpful discussions and suggestions. I am also grateful to Jan Star\'{y} for language corrections.
The 3D laser scanning was performed with the financial support of the Praemium Academicae of Academy of Sciences of the Czech Republic. 
\end{acknowledgements}

\bibliography{cmri}

\end{document}